\documentclass[aps,twocolumn,showpacs,preprintnumbers,amsmath,amssymb,superscriptaddress,longbibliography]{revtex4-2}
\usepackage{graphicx}
\usepackage{dcolumn}
\usepackage{bm}
\usepackage{braket}
\usepackage{amsmath}
\usepackage[normalem]{ulem}

\begin{document}

\title{Tan's Contact for Trapped Spin-Orbit-Coupled Fermi Gases at Finite Temperature}

\author{Tianhao Wu}
\affiliation{Ministry of Education Key Laboratory for Nonequilibrium Synthesis and Modulation of Condensed Matter,Shaanxi Province Key Laboratory of Quantum Information and Quantum Optoelectronic Devices, School of Physics, Xi'an Jiaotong University, Xi'an 710049, China}

\author{Yuanbang Wei}
\affiliation{Ministry of Education Key Laboratory for Nonequilibrium Synthesis and Modulation of Condensed Matter,Shaanxi Province Key Laboratory of Quantum Information and Quantum Optoelectronic Devices, School of Physics, Xi'an Jiaotong University, Xi'an 710049, China}

\author{Rui Tian}
\affiliation{Ministry of Education Key Laboratory for Nonequilibrium Synthesis and Modulation of Condensed Matter,Shaanxi Province Key Laboratory of Quantum Information and Quantum Optoelectronic Devices, School of Physics, Xi'an Jiaotong University, Xi'an 710049, China}

\author{Dongdong Pu}
\affiliation{Ministry of Education Key Laboratory for Nonequilibrium Synthesis and Modulation of Condensed Matter,Shaanxi Province Key Laboratory of Quantum Information and Quantum Optoelectronic Devices, School of Physics, Xi'an Jiaotong University, Xi'an 710049, China}

\author{Beizhi Chen}
\affiliation{Ministry of Education Key Laboratory for Nonequilibrium Synthesis and Modulation of Condensed Matter,Shaanxi Province Key Laboratory of Quantum Information and Quantum Optoelectronic Devices, School of Physics, Xi'an Jiaotong University, Xi'an 710049, China}

\author{Qihang Ying}
\affiliation{Ministry of Education Key Laboratory for Nonequilibrium Synthesis and Modulation of Condensed Matter,Shaanxi Province Key Laboratory of Quantum Information and Quantum Optoelectronic Devices, School of Physics, Xi'an Jiaotong University, Xi'an 710049, China}

\author{Hongyan Shi}
\affiliation{Ministry of Education Key Laboratory for Nonequilibrium Synthesis and Modulation of Condensed Matter,Shaanxi Province Key Laboratory of Quantum Information and Quantum Optoelectronic Devices, School of Physics, Xi'an Jiaotong University, Xi'an 710049, China}

\author{Yue Zhang}
\affiliation{Ministry of Education Key Laboratory for Nonequilibrium Synthesis and Modulation of Condensed Matter,Shaanxi Province Key Laboratory of Quantum Information and Quantum Optoelectronic Devices, School of Physics, Xi'an Jiaotong University, Xi'an 710049, China}

\author{Shuai Li}
\email{lishuai0999@xjtu.edu.cn}
\affiliation{Ministry of Education Key Laboratory for Nonequilibrium Synthesis and Modulation of Condensed Matter,Shaanxi Province Key Laboratory of Quantum Information and Quantum Optoelectronic Devices, School of Physics, Xi'an Jiaotong University, Xi'an 710049, China}

\author{Bo Liu}
\email{liubophy@xjtu.edu.cn}
\affiliation{Ministry of Education Key Laboratory for Nonequilibrium Synthesis and Modulation of Condensed Matter,Shaanxi Province Key Laboratory of Quantum Information and Quantum Optoelectronic Devices, School of Physics, Xi'an Jiaotong University, Xi'an 710049, China}

\begin{abstract}
Tan's contact, bridging microscopic features of many-body quantum systems to two-body contact interactions, is pivotal in the description of quantum gases. However, past studies of Tan's contact of the spin-orbit coupled Fermi gases have been exclusively discussed at zero temperature. The thermal effects on that, which cannot be neglected in most experimental conditions, has not been explored so far. Here, through a path integral Monte Carlo (PIMC) study, we show that thermal effects fundamentally alter the scaling relation between spin-orbit coupling (SOC) and Tan's contact. The linear scaling of Tan's contact with respect to the strength of SOC found at zero temperature is drastically undermined by the thermal effects. And a new mechanism based on the competition between the mean occupancy and the energy-reweighted occupancy of the opposite spin fermion pairs within the interaction range has been unveiled, which radically changes the above linear scaling to a non-monotonic dependence. Our work should be useful for future experiments to explore the preternatural thermal effects on Tan's contact of the strongly interacting spin-orbit-coupled Fermi gas.
\end{abstract}

\maketitle

Understanding strongly interacting fermions is one of the most challenging problems in
modern physics. Strongly interacting fermions have wide-ranging
implications for systems such as quark matter in neutron stars \cite{lattimer2004physics} and high-temperature
superconductors \cite{giorgini2008theory, lee2006doping}. However, elucidating the behavior of interacting fermions in the correlated strongly interacting regime remains elusive. An important generic idea in this field is the fermionic universality: all
strongly interacting, dilute Fermi gases should behave identically, depending only on a scaling
factor equal to the average particle separation, but not on the details of the interaction \cite{ho2004universal}.

The realization of broad collisional (Feshbach) resonances with an interaction
potential range being small relative to the inter-particle spacing in ultracold atomic Fermi
gases provides a highly controlled way for studying strongly
interacting fermions \cite{chin2010feshbach, ohara2002observation, bloch2008many}.
One of the key quantities characterizing the universal properties of the strongly interacting Fermi gases is the integrated contact intensity or ``contact'' $C$, introduced by Tan \cite{tan2008energetics, tan2008large, tan2008generalized}. The contact derives from the short-range correlations in strongly interacting quantum gases and is the cornerstone of a number of exact relations describing properties such as the equation of state and dynamic response functions \cite{braaten2008exact}. Recent experiments utilizing techniques such as ballistic expansion, Radio-Frequency Spectroscopy, and atom photoemission spectroscopy have successfully verified Tan's relations in various quantum gas setups \cite{stewart2010verification, kuhnle2010universal, navon2010equation}.

On the other hand, the experimental realization of synthetic spin-orbit
coupling (SOC) in ultracold atomic gases greatly extends
their tunability \cite{lin2011spin, wang2012spin, cheuk2012spin}. SOC can significantly impact both single-particle and
many-body properties \cite{galitski2013spin, zhai2015degenerate}. So far, past studies of the contact theory for spin-orbit-coupled
systems have exclusively focused on the perturbation context at zero temperature \cite{Peng18,Peng20}.
In most experimental conditions, thermal effects of the gases cannot be neglected \cite{wang2012spin,fu2013radio}.
However, the thermal effects on the contact of spin-orbit-coupled strongly interacting Fermi gases have remained unexplored.

In this work, we first uncover the significant thermal effects on the contact of spin-orbit-coupled strongly interacting Fermi gases by performing a path integral Monte Carlo (PIMC) study \cite{PIMC95,ceperley1991fermion,PIMC05,boninsegni2006worm,PIMC13,PIMC14,dornheim2018uniform}. Explicitly, we show that: (i) thermal effects drastically undermine the linear scaling of the Tan's contact with respect to the strength of SOC found at zero temperature and dramatically shrink the linear-dependent region; (ii) thermal effects radically change the linear scaling of the Tan's contact with respect to the strength of SOC to a non-monotonic dependence. The underlying mechanism is also unveiled, which is due to the competition between the mean occupancy and the energy-reweighted occupancy of the opposite-spin fermion pairs within the interaction range. Our work thus paves the way for exploring the profound thermal effects on strongly interacting spin-orbit-coupled Fermi gases.

\begin{figure}[t]
	\centering
	\includegraphics[width=0.49\textwidth]{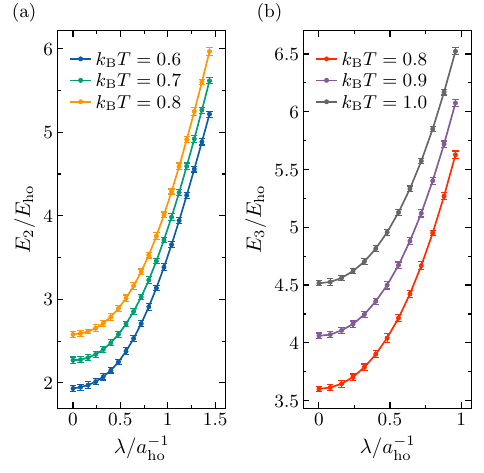}
\caption{Total energy $E$ as a function of SOC strength $\lambda$ for different temperature and total particle number. (a) Total energy $E_2$ for two-particle system. Here we choose three different temperatures, i.e., $k_B T/E_{\text{ho}} = 0.6$, $0.7$, and $0.8$, where $E_{\text{ho}}=\hbar\omega$. (b) Total energy $E_3$ for three-particle system, fixing temperature at $k_B T/E_{\text{ho}} = 0.8$, $0.9$, and $1.0$, respectively.}
\label{fig2}
\end{figure}

\emph{PIMC study of a SOC Fermi gas}$\raisebox{0.01mm}{---}$ Let us consider a spin-orbit-coupled Fermi gas consisting of $N$ atoms with
mass $m$, where $\mathbf{r}_j$ $(j = 1,\cdots,N)$ describes the corresponding positions of these $N$ fermions. The system can be captured by the following model Hamiltonian ${H}$,
\begin{equation}
\begin{split}
{H} &=\sum_{j=1}^N \left[ -\frac{\hbar^2}{2m}\nabla^2_j+\frac{m\omega^2\mathbf{r}_j^2}{2}+\frac{\hbar^2\lambda}{m}\left(-i\nabla_j\right)\cdot\hat{\boldsymbol{\sigma}}\right]
\\ &\quad + \sum_{j<k}
    V_{j,k}\delta_{\sigma_j,-\sigma_k}+\frac{N\hbar^2\lambda^2}{2m} ,
\end{split}
\label{eq:H_total}
\end{equation}
where $\hat{\boldsymbol{\sigma}}=(\hat{\sigma}_x, \hat{\sigma}_y, \hat{\sigma}_z)$ is the Pauli vector and $\lambda$ represents the SOC strength.
$\omega$ is the frequency of the harmonic trapping potential.
The interaction between fermions with opposite spins labeled by $\sigma$ and $-\sigma$ is modeled by a square-well potential, which can be expressed as
\begin{equation}
    V_{j,k} = 
    \begin{cases}
        V_0 & r_{jk} < r_0 \\
        0 & r_{jk} \ge r_0
    \end{cases},
    \label{eq:V_sw}
\end{equation}
with $r_{jk}=|\mathbf{r}_j - \mathbf{r}_k|$ denoting the distance between two particles. $V_0 < 0$ is the potential depth, and $r_0$ is its range.

In the following, we employ the path integral Monte Carlo (PIMC) method to study the model system in Eq.~(\ref{eq:H_total}). We discretize imaginary time $\beta = 1/k_B T$ into $M$ time steps, namely $\tau = \beta/M$, and define $\mathbf{x} \equiv \{ \mathbf{R}_0, \dots, \mathbf{R}_{M-1} \}$ \cite{PIMC95,PIMC13,PIMC14}, where $\mathbf{R}_j$ represents both spatial and spin coordinates of all $N$ particles at $j$-th time step. The many-particle partition function $Z = \text{Tr}(e^{-\beta \hat{H}})$ can be expressed as
\begin{equation}
    Z = \frac{1}{N!} \sum_{\mathcal{P}} (-1)^{\mathcal{P}} \int d\mathbf{x} \prod_{j=0}^{M-1} \rho(\mathbf{R}_j, \mathbf{R}_{j+1}; \tau).
    \label{eq:Z_pimc}
\end{equation}
The path closure in imaginary time is enforced by imposing the boundary condition $\mathbf{R}_{M} = \mathcal{P}\mathbf{R}_0$, and $(-1)^{\mathcal{P}}$ captures the sign due to permutation of fermions.

For a sufficiently small time step $\tau$, we employ the Trotter-Suzuki decomposition \cite{PIMC95,PIMC05} to factorize the density matrix $\rho$ as
\begin{equation}
    \rho(\mathbf{R}, \mathbf{R}'; \tau) = A_F(\mathbf{R}, \mathbf{R}'; \tau) \prod_{j<k} \exp\Bigl( -\tau V_{j,k}\delta_{\sigma_j,-\sigma_k} \Bigr),
    \label{eq:rho_factorized}
\end{equation}
where $A_F(\mathbf{R}, \mathbf{R}'; \tau)=\prod_{j=1}^N\rho_{F}(\mathbf{r}_j, \mathbf{r}'_j,\sigma_{j},\sigma'_{j}; \tau)$. Here $\rho_{F}$ can be approximated in the short-time expansion \cite{PIMC95,PIMC05} as
\begin{equation}
\begin{split}
    \rho_{F} &= \left(\frac{m}{2\pi\tau}\right)^{\frac{3}{2}}\exp\left(\frac{\tau\lambda^2}{m}-\frac{m|\Delta\mathbf{r}_{j}|^2}{2\tau}\right)
    \\ &\quad \times\exp\left(-\frac{\tau m\omega^2}{2}\mathbf{r}_{j}^2\right)\langle \sigma'_{j} | U | \sigma_{j} \rangle,
\end{split}
\label{eq:rho_single}
\end{equation}
with $\Delta\mathbf{r}_{j}=\mathbf{r}'_j-\mathbf{r}_j$ and $U$ capturing the spin part in Eq.~(\ref{eq:rho_single}) can be expressed as
\begin{equation}
    U = \cos(\lambda |\Delta\mathbf{r}_{j}|) \mathbf{I} - i \sin(\lambda |\Delta\mathbf{r}_{j}|) \frac{\Delta\mathbf{r}_{j} \cdot \hat{\boldsymbol{\sigma}}}{|\Delta\mathbf{r}_{j}|},
\end{equation}
where $\mathbf{I}$ is the identity matrix.



\emph{Thermal effects  on Tan's contact}$\raisebox{0.01mm}{---}$ To compute the Tan's contact \cite{tan2008energetics, tan2008large, tan2008generalized}, defined as $C/(4\pi m)=\partial E/\partial(-a_s^{-1})$, in our PIMC simulations, we obtain the thermodynamic estimator of the energy \cite{PIMC95} through the following relation $E=-\partial\ln Z/\partial\beta$. Here we adjust $V_0$ such that the s-wave scattering length $a_s$ diverges \cite{Cui12}. $r_0$ is fixed at $r_0=0.1a_{\mathrm{ho}}$ with $a_{\mathrm{ho}}=\sqrt{\hbar/(m\omega)}$. As shown in Fig.~\ref{fig2}, $E$ is computed from our PIMC simulations for different temperature and total particle number.

\begin{figure*}[t]
	\centering
	\includegraphics[width=1.0\textwidth]{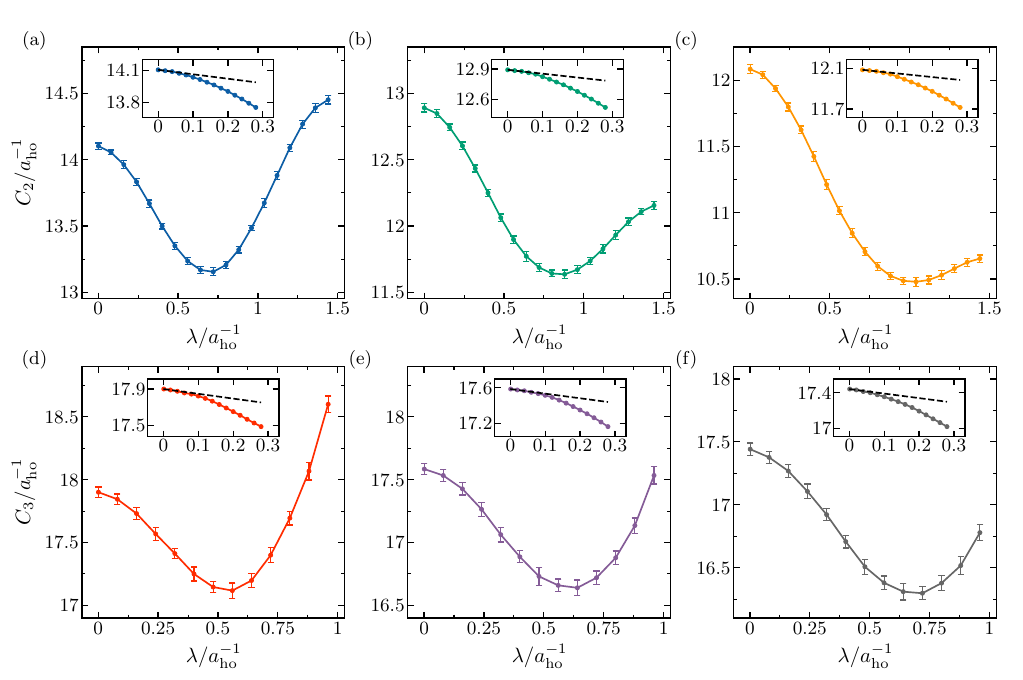}
\caption{Tan's contact as the function of SOC strength $\lambda$. (a--c) show the Tan's contact for two-particle system, while (d--f) display results for three-particle system, which are obtained using the PIMC approach. Here we choose different temperatures, i.e., $k_B T/E_{\text{ho}} = 0.6$, $0.7$, $0.8$, $0.8$, $0.9$, and $1.0$ for (a--f), respectively. The insets show the results in the weak spin-orbit coupled regime, and the dash line stands for the linear dependence.}
\label{fig1}
\end{figure*}

\begin{figure}[t]
	\centering
	\includegraphics[width=0.49\textwidth]{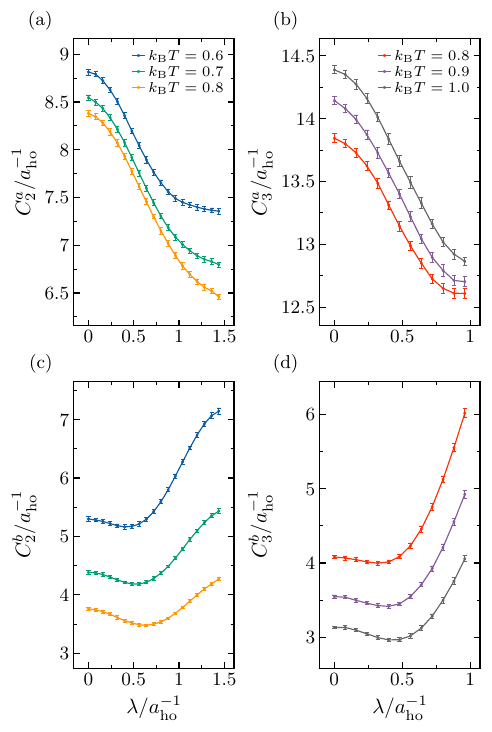}
\caption{$C^a$ and $C^b$ defined in Tan's contact $C$ as the function of SOC strength $\lambda$. (a) and (c) $C^a_2$ and $C^b_2$ for two-particle system, we choose temperatures at $k_B T/E_{\text{ho}} = 0.6$, $0.7$, and $0.8$, respectively. (b) and (d) $C^a_3$ and $C^b_3$ for three-particle system when choosing temperatures at $k_B T/E_{\text{ho}} = 0.8$, $0.9$, and $1.0$, respectively.}
\label{fig3}
\end{figure}

We then further rewrite the above contact relation as 
\begin{equation}
\begin{split}
\frac{\partial E}{\partial (-a_s^{-1})} &=\frac{\partial Z}{Z\partial\beta}\frac{\partial Z}{Z\partial (-a_s^{-1})}-\frac{\partial\frac{\partial Z}{\partial\beta}}{Z\partial (-a_s^{-1})}
\\ &= \left\langle \frac{\partial\ln\pi}{\partial (-a_s^{-1})} \right\rangle \left\langle \frac{\partial\ln\pi}{\partial\beta} \right\rangle \\
&\quad - \left\langle \frac{\partial\ln\pi}{\partial(-a_s^{-1})}\frac{\partial\ln\pi}{\partial\beta}\right\rangle  - \left\langle \frac{\partial^2\ln\pi}{\partial(-a_s^{-1})\partial\beta} \right\rangle \\
&= \frac{C^a}{4\pi m}+\frac{C^b}{4\pi m},
\end{split}
\label{eq:contact_estimator_raw}
\end{equation}
where $\pi(\mathbf{x}) = \prod_{j=0}^{M-1} \rho(\mathbf{R}_j, \mathbf{R}_{j+1}; \tau)$, $C^a/(4\pi m) := -\langle\frac{\partial^2\ln\pi}{\partial(-a_s^{-1})\partial\beta}\rangle$ and $C^b/(4\pi m) :=\langle\frac{\partial\ln\pi}{\partial(-a_s^{-1})}\rangle\langle\frac{\partial\ln\pi}{\partial\beta}\rangle-\langle\frac{\partial\ln\pi}{\partial(-a_s^{-1})}\frac{\partial\ln\pi}{\partial\beta}\rangle$.
Through obtaining all the thermal expectation values in the right hand side of Eq.~(\ref{eq:contact_estimator_raw}) from our PIMC simulation, we can obtain the Tan's contact $C$.




In most experimental conditions, thermal effects cannot be neglected. In our PIMC study, we unveil that such thermal effects can substantially change the Tan's contact, which can characterize many-body quantum systems, bridging microscopic correlations to macroscopic thermodynamic behavior.
First, we find that thermal effects can drastically undermine the linear scaling of the Tan's contact with respect to the strength of SOC found at zero temperature \cite{Peng18,Peng20}. As shown in Fig.~\ref{fig1}, our PIMC results show that thermal effects drastically shrink the region where the linear scaling of the Tan's contact with respect to the strength of SOC exists. The linear scaling is only valid when considering extremely weak spin-orbit coupling.

Second, more importantly, our PIMC studies show that thermal effects will change the linear scaling of the Tan's contact with respect to the strength of SOC to a non-monotonic dependence. To understand that, we define the time-slice-averaged number of opposite-spin fermion pairs within the interaction range as

\begin{equation}
\bar n(\mathbf{x})=\frac{1}{M}\sum_{\ell=0}^{M-1}\sum_{i<k}
\Theta(r_0-r_{\ell,ik})\delta_{\sigma_{\ell,i},-\sigma_{\ell,k}},
\label{eq:pair_count_physical}
\end{equation}
where $r_{\ell,ik}$ is the pair separation on slice $\ell$ and $\Theta(r_0-r)$ stands for a step function. Then, we can rewrite $C^a$ and $C^b$ in Eq.~(\ref{eq:contact_estimator_raw}) as  
\begin{align}
\frac{C^a}{4\pi m}&=\frac{dV_0}{d(-a_s^{-1})}\langle\bar n\rangle,\notag\\
\frac{C^b}{4\pi m}&=-\beta \frac{dV_0}{d(-a_s^{-1})}\operatorname{Cov}(\mathcal E,\bar n),
\label{eq:contact_pair_physical}
\end{align}
where $\operatorname{Cov}(\mathcal E,\bar n)=\langle\mathcal E\bar n\rangle-\langle\mathcal E\rangle\langle\bar n\rangle$ and the
energy estimator $\mathcal E=-\partial_\beta\ln\pi$.

From Eq.~(\ref{eq:contact_pair_physical}), it is shown that $C^a$ describes the mean occupancy of opposite-spin fermion pairs, which decreases monotonically with increasing the SOC strength, for instance, as shown in Fig.~\ref{fig3}.
Regarding $C^b$, to understand its underlying physics, we further rewrite it as
\begin{equation}
\frac{C^b}{4\pi m}=-\beta \frac{dV_0}{d(-a_s^{-1})}\langle\mathcal E\rangle
\bigl(\langle\bar n\rangle_{\mathcal E}-\langle\bar n\rangle\bigr),
\end{equation}
via defining the normalized energy-reweighted average as $\langle\bar n\rangle_{\mathcal E}
=\frac{\langle\mathcal E\bar n\rangle}{\langle\mathcal E\rangle}$.
Then, we can find that $C^b$ describes the change of the mean pair occupancy under the energy reweighting. As shown in Fig.~\ref{fig3}, $C^b$ initially decreases and then increases when enhancing the strength of SOC.

The non-monotonic dependence of $C$ on $\lambda$ consequently reflects the competition between $C^a$ and $C^b$. In the weak SOC region, both $C^a$ and $C^b$ monotonically decrease when increasing strength of the SOC. When further enhancing the SOC strength, $C^b$ will increase, while $C^a$ still keeps decreasing. The flex point of $C$ is determined via the relation $\partial_\lambda C^a+\partial_\lambda C^b=0$. Therefore, the competition between the mean occupancy and the energy-reweighted occupancy of the opposite spin fermion pairs within the interaction range results in the non-monotonic scaling of $C$ with respect to the strength of SOC.

\textit{Discussion} \& \textit{Conclusion}$\raisebox{0.01mm}{---}$ In summary, we have investigated the thermal effects on Tan's contact of a strongly interacting spin-orbit-coupled Fermi gas implementing a path integral Monte Carlo (PIMC) study. Our work extends the understanding of Tan's contact beyond the past studies focusing on the zero-temperature regime, revealing that finite-temperature effects fundamentally alter the properties of strongly interacting fermions with SOC.

First, we demonstrated that the linear scaling of the Tan's contact with respect to the strength of SOC found at zero temperature is severely constrained by thermal effects. The valid regime for the linear SOC dependence of Tan's contact is dramatically shrunk.
Most importantly, we uncovered a non-monotonic dependence of Tan's contact on the SOC strength and identified the underlying physics of its two components. One refers to the mean occupancy of opposite-spin fermion pairs within the interaction range, and the other captures the opposite-spin fermion pairs contribution from the redistribution of path weights. Due to the competition between them, the new non-monotonic dependence appears. Our work should be useful for future experiments to explore the preternatural thermal effects on Tan's contact of the strongly interacting spin-orbit-coupled Fermi gas. 


\textit{Acknowledgements$\raisebox{0.01mm}{---}$}This work is supported by the National Key R$\&$D Program of China (2021YFA1401700), NSFC (Grants No. 12474267, 12074305), the Fundamental Research Funds for the Central Universities (Grant No. xtr052023002), the Shaanxi Fundamental Science Research Project for Mathematics and Physics (Grant No. 23JSZ003), Shanghai Municipal Science and Technology Major Project (Grant No. 2019SHZDZX01), and Natural Science Basic Research Program of Shaanxi (Grant No. 2025JC-YBQN-897).

\bibliographystyle{apsrev}
\bibliography{soc}

\begin{thebibliography}{30}
\expandafter\ifx\csname natexlab\endcsname\relax\def\natexlab#1{#1}\fi
\expandafter\ifx\csname bibnamefont\endcsname\relax
  \def\bibnamefont#1{#1}\fi
\expandafter\ifx\csname bibfnamefont\endcsname\relax
  \def\bibfnamefont#1{#1}\fi
\expandafter\ifx\csname citenamefont\endcsname\relax
  \def\citenamefont#1{#1}\fi
\expandafter\ifx\csname url\endcsname\relax
  \def\url#1{\texttt{#1}}\fi
\expandafter\ifx\csname urlprefix\endcsname\relax\def\urlprefix{URL }\fi
\providecommand{\bibinfo}[2]{#2}
\providecommand{\eprint}[2][]{\url{#2}}

\bibitem[{\citenamefont{Lattimer and Prakash}(2004)}]{lattimer2004physics}
\bibinfo{author}{\bibfnamefont{J.~M.} \bibnamefont{Lattimer}} \bibnamefont{and}
  \bibinfo{author}{\bibfnamefont{M.}~\bibnamefont{Prakash}},
  \bibinfo{journal}{Science} \textbf{\bibinfo{volume}{304}},
  \bibinfo{pages}{536} (\bibinfo{year}{2004}).

\bibitem[{\citenamefont{Giorgini et~al.}(2008)\citenamefont{Giorgini,
  Pitaevskii, and Stringari}}]{giorgini2008theory}
\bibinfo{author}{\bibfnamefont{S.}~\bibnamefont{Giorgini}},
  \bibinfo{author}{\bibfnamefont{L.~P.} \bibnamefont{Pitaevskii}},
  \bibnamefont{and}
  \bibinfo{author}{\bibfnamefont{S.}~\bibnamefont{Stringari}},
  \bibinfo{journal}{Rev. Mod. Phys.} \textbf{\bibinfo{volume}{80}},
  \bibinfo{pages}{1215} (\bibinfo{year}{2008}).

\bibitem[{\citenamefont{Lee et~al.}(2006)\citenamefont{Lee, Nagaosa, and
  Wen}}]{lee2006doping}
\bibinfo{author}{\bibfnamefont{P.~A.} \bibnamefont{Lee}},
  \bibinfo{author}{\bibfnamefont{N.}~\bibnamefont{Nagaosa}}, \bibnamefont{and}
  \bibinfo{author}{\bibfnamefont{X.-G.} \bibnamefont{Wen}},
  \bibinfo{journal}{Rev. Mod. Phys.} \textbf{\bibinfo{volume}{78}},
  \bibinfo{pages}{17} (\bibinfo{year}{2006}).

\bibitem[{\citenamefont{Ho}(2004)}]{ho2004universal}
\bibinfo{author}{\bibfnamefont{T.-L.} \bibnamefont{Ho}},
  \bibinfo{journal}{Phys. Rev. Lett.} \textbf{\bibinfo{volume}{92}},
  \bibinfo{pages}{090402} (\bibinfo{year}{2004}).

\bibitem[{\citenamefont{Chin et~al.}(2010)\citenamefont{Chin, Grimm, Julienne,
  and Tiesinga}}]{chin2010feshbach}
\bibinfo{author}{\bibfnamefont{C.}~\bibnamefont{Chin}},
  \bibinfo{author}{\bibfnamefont{R.}~\bibnamefont{Grimm}},
  \bibinfo{author}{\bibfnamefont{P.}~\bibnamefont{Julienne}}, \bibnamefont{and}
  \bibinfo{author}{\bibfnamefont{E.}~\bibnamefont{Tiesinga}},
  \bibinfo{journal}{Rev. Mod. Phys.} \textbf{\bibinfo{volume}{82}},
  \bibinfo{pages}{1225} (\bibinfo{year}{2010}).

\bibitem[{\citenamefont{O'Hara et~al.}(2002)\citenamefont{O'Hara, Hemmer, Gehm,
  Granade, and Thomas}}]{ohara2002observation}
\bibinfo{author}{\bibfnamefont{K.~M.} \bibnamefont{O'Hara}},
  \bibinfo{author}{\bibfnamefont{S.~L.} \bibnamefont{Hemmer}},
  \bibinfo{author}{\bibfnamefont{M.~E.} \bibnamefont{Gehm}},
  \bibinfo{author}{\bibfnamefont{S.~R.} \bibnamefont{Granade}},
  \bibnamefont{and} \bibinfo{author}{\bibfnamefont{J.~E.}
  \bibnamefont{Thomas}}, \bibinfo{journal}{Science}
  \textbf{\bibinfo{volume}{298}}, \bibinfo{pages}{2179} (\bibinfo{year}{2002}).

\bibitem[{\citenamefont{Bloch et~al.}(2008)\citenamefont{Bloch, Dalibard, and
  Zwerger}}]{bloch2008many}
\bibinfo{author}{\bibfnamefont{I.}~\bibnamefont{Bloch}},
  \bibinfo{author}{\bibfnamefont{J.}~\bibnamefont{Dalibard}}, \bibnamefont{and}
  \bibinfo{author}{\bibfnamefont{W.}~\bibnamefont{Zwerger}},
  \bibinfo{journal}{Rev. Mod. Phys.} \textbf{\bibinfo{volume}{80}},
  \bibinfo{pages}{885} (\bibinfo{year}{2008}).

\bibitem[{\citenamefont{Tan}(2008{\natexlab{a}})}]{tan2008energetics}
\bibinfo{author}{\bibfnamefont{S.}~\bibnamefont{Tan}}, \bibinfo{journal}{Ann.
  Phys.} \textbf{\bibinfo{volume}{323}}, \bibinfo{pages}{2952}
  (\bibinfo{year}{2008}{\natexlab{a}}).

\bibitem[{\citenamefont{Tan}(2008{\natexlab{b}})}]{tan2008large}
\bibinfo{author}{\bibfnamefont{S.}~\bibnamefont{Tan}}, \bibinfo{journal}{Ann.
  Phys.} \textbf{\bibinfo{volume}{323}}, \bibinfo{pages}{2971}
  (\bibinfo{year}{2008}{\natexlab{b}}).

\bibitem[{\citenamefont{Tan}(2008{\natexlab{c}})}]{tan2008generalized}
\bibinfo{author}{\bibfnamefont{S.}~\bibnamefont{Tan}}, \bibinfo{journal}{Ann.
  Phys.} \textbf{\bibinfo{volume}{323}}, \bibinfo{pages}{2987}
  (\bibinfo{year}{2008}{\natexlab{c}}).

\bibitem[{\citenamefont{Braaten and Platter}(2008)}]{braaten2008exact}
\bibinfo{author}{\bibfnamefont{E.}~\bibnamefont{Braaten}} \bibnamefont{and}
  \bibinfo{author}{\bibfnamefont{L.}~\bibnamefont{Platter}},
  \bibinfo{journal}{Phys. Rev. Lett.} \textbf{\bibinfo{volume}{100}},
  \bibinfo{pages}{205301} (\bibinfo{year}{2008}).

\bibitem[{\citenamefont{Stewart et~al.}(2010)\citenamefont{Stewart, Gaebler,
  Drake, and Jin}}]{stewart2010verification}
\bibinfo{author}{\bibfnamefont{J.~T.} \bibnamefont{Stewart}},
  \bibinfo{author}{\bibfnamefont{J.~P.} \bibnamefont{Gaebler}},
  \bibinfo{author}{\bibfnamefont{T.~E.} \bibnamefont{Drake}}, \bibnamefont{and}
  \bibinfo{author}{\bibfnamefont{D.~S.} \bibnamefont{Jin}},
  \bibinfo{journal}{Phys. Rev. Lett.} \textbf{\bibinfo{volume}{104}},
  \bibinfo{pages}{235301} (\bibinfo{year}{2010}).

\bibitem[{\citenamefont{Kuhnle et~al.}(2010)\citenamefont{Kuhnle, Hu, Liu,
  Dyke, Mark, Drummond, Hannaford, and Vale}}]{kuhnle2010universal}
\bibinfo{author}{\bibfnamefont{E.~D.} \bibnamefont{Kuhnle}},
  \bibinfo{author}{\bibfnamefont{H.}~\bibnamefont{Hu}},
  \bibinfo{author}{\bibfnamefont{X.-J.} \bibnamefont{Liu}},
  \bibinfo{author}{\bibfnamefont{P.}~\bibnamefont{Dyke}},
  \bibinfo{author}{\bibfnamefont{M.}~\bibnamefont{Mark}},
  \bibinfo{author}{\bibfnamefont{P.~D.} \bibnamefont{Drummond}},
  \bibinfo{author}{\bibfnamefont{P.}~\bibnamefont{Hannaford}},
  \bibnamefont{and} \bibinfo{author}{\bibfnamefont{C.~J.} \bibnamefont{Vale}},
  \bibinfo{journal}{Phys. Rev. Lett.} \textbf{\bibinfo{volume}{105}},
  \bibinfo{pages}{070402} (\bibinfo{year}{2010}).

\bibitem[{\citenamefont{Navon et~al.}(2010)\citenamefont{Navon, Nascimb{\`e}ne,
  Chevy, and Salomon}}]{navon2010equation}
\bibinfo{author}{\bibfnamefont{N.}~\bibnamefont{Navon}},
  \bibinfo{author}{\bibfnamefont{S.}~\bibnamefont{Nascimb{\`e}ne}},
  \bibinfo{author}{\bibfnamefont{F.}~\bibnamefont{Chevy}}, \bibnamefont{and}
  \bibinfo{author}{\bibfnamefont{C.}~\bibnamefont{Salomon}},
  \bibinfo{journal}{Science} \textbf{\bibinfo{volume}{328}},
  \bibinfo{pages}{729} (\bibinfo{year}{2010}).

\bibitem[{\citenamefont{Lin et~al.}(2011)\citenamefont{Lin,
  Jim{\'e}nez-Garc{\'\i}a, and Spielman}}]{lin2011spin}
\bibinfo{author}{\bibfnamefont{Y.-J.} \bibnamefont{Lin}},
  \bibinfo{author}{\bibfnamefont{K.}~\bibnamefont{Jim{\'e}nez-Garc{\'\i}a}},
  \bibnamefont{and} \bibinfo{author}{\bibfnamefont{I.~B.}
  \bibnamefont{Spielman}}, \bibinfo{journal}{Nature}
  \textbf{\bibinfo{volume}{471}}, \bibinfo{pages}{83} (\bibinfo{year}{2011}).

\bibitem[{\citenamefont{Wang et~al.}(2012)\citenamefont{Wang, Yu, Fu, Miao,
  Huang, Chai, Zhai, and Zhang}}]{wang2012spin}
\bibinfo{author}{\bibfnamefont{P.}~\bibnamefont{Wang}},
  \bibinfo{author}{\bibfnamefont{Z.-Q.} \bibnamefont{Yu}},
  \bibinfo{author}{\bibfnamefont{Z.}~\bibnamefont{Fu}},
  \bibinfo{author}{\bibfnamefont{J.}~\bibnamefont{Miao}},
  \bibinfo{author}{\bibfnamefont{L.}~\bibnamefont{Huang}},
  \bibinfo{author}{\bibfnamefont{S.}~\bibnamefont{Chai}},
  \bibinfo{author}{\bibfnamefont{H.}~\bibnamefont{Zhai}}, \bibnamefont{and}
  \bibinfo{author}{\bibfnamefont{J.}~\bibnamefont{Zhang}},
  \bibinfo{journal}{Phys. Rev. Lett.} \textbf{\bibinfo{volume}{109}},
  \bibinfo{pages}{095301} (\bibinfo{year}{2012}).

\bibitem[{\citenamefont{Cheuk et~al.}(2012)\citenamefont{Cheuk, Sommer,
  Hadzibabic, Yefsah, Bakr, and Zwierlein}}]{cheuk2012spin}
\bibinfo{author}{\bibfnamefont{L.~W.} \bibnamefont{Cheuk}},
  \bibinfo{author}{\bibfnamefont{A.~T.} \bibnamefont{Sommer}},
  \bibinfo{author}{\bibfnamefont{Z.}~\bibnamefont{Hadzibabic}},
  \bibinfo{author}{\bibfnamefont{T.}~\bibnamefont{Yefsah}},
  \bibinfo{author}{\bibfnamefont{W.~S.} \bibnamefont{Bakr}}, \bibnamefont{and}
  \bibinfo{author}{\bibfnamefont{M.~W.} \bibnamefont{Zwierlein}},
  \bibinfo{journal}{Phys. Rev. Lett.} \textbf{\bibinfo{volume}{109}},
  \bibinfo{pages}{095302} (\bibinfo{year}{2012}).

\bibitem[{\citenamefont{Galitski and Spielman}(2013)}]{galitski2013spin}
\bibinfo{author}{\bibfnamefont{V.}~\bibnamefont{Galitski}} \bibnamefont{and}
  \bibinfo{author}{\bibfnamefont{I.~B.} \bibnamefont{Spielman}},
  \bibinfo{journal}{Nature} \textbf{\bibinfo{volume}{494}}, \bibinfo{pages}{49}
  (\bibinfo{year}{2013}).

\bibitem[{\citenamefont{Zhai}(2015)}]{zhai2015degenerate}
\bibinfo{author}{\bibfnamefont{H.}~\bibnamefont{Zhai}}, \bibinfo{journal}{Rep.
  Prog. Phys.} \textbf{\bibinfo{volume}{78}}, \bibinfo{pages}{026001}
  (\bibinfo{year}{2015}).

\bibitem[{\citenamefont{Peng et~al.}(2018)\citenamefont{Peng, Zhang, Tan, and
  Jiang}}]{Peng18}
\bibinfo{author}{\bibfnamefont{S.-G.} \bibnamefont{Peng}},
  \bibinfo{author}{\bibfnamefont{C.-X.} \bibnamefont{Zhang}},
  \bibinfo{author}{\bibfnamefont{S.}~\bibnamefont{Tan}}, \bibnamefont{and}
  \bibinfo{author}{\bibfnamefont{K.}~\bibnamefont{Jiang}},
  \bibinfo{journal}{Phys. Rev. Lett.} \textbf{\bibinfo{volume}{120}},
  \bibinfo{pages}{060408} (\bibinfo{year}{2018}).

\bibitem[{\citenamefont{Zhang et~al.}(2020)\citenamefont{Zhang, Peng, and
  Jiang}}]{Peng20}
\bibinfo{author}{\bibfnamefont{C.-X.} \bibnamefont{Zhang}},
  \bibinfo{author}{\bibfnamefont{S.-G.} \bibnamefont{Peng}}, \bibnamefont{and}
  \bibinfo{author}{\bibfnamefont{K.}~\bibnamefont{Jiang}},
  \bibinfo{journal}{Phys. Rev. A} \textbf{\bibinfo{volume}{101}},
  \bibinfo{pages}{043616} (\bibinfo{year}{2020}).

\bibitem[{\citenamefont{Fu et~al.}(2013)\citenamefont{Fu, Huang, Meng, Wang,
  Liu, Pu, Hu, and Zhang}}]{fu2013radio}
\bibinfo{author}{\bibfnamefont{Z.}~\bibnamefont{Fu}},
  \bibinfo{author}{\bibfnamefont{L.}~\bibnamefont{Huang}},
  \bibinfo{author}{\bibfnamefont{Z.}~\bibnamefont{Meng}},
  \bibinfo{author}{\bibfnamefont{P.}~\bibnamefont{Wang}},
  \bibinfo{author}{\bibfnamefont{X.-J.} \bibnamefont{Liu}},
  \bibinfo{author}{\bibfnamefont{H.}~\bibnamefont{Pu}},
  \bibinfo{author}{\bibfnamefont{H.}~\bibnamefont{Hu}}, \bibnamefont{and}
  \bibinfo{author}{\bibfnamefont{J.}~\bibnamefont{Zhang}},
  \bibinfo{journal}{Phys. Rev. A} \textbf{\bibinfo{volume}{87}},
  \bibinfo{pages}{053619} (\bibinfo{year}{2013}).

\bibitem[{\citenamefont{Ceperley}(1995)}]{PIMC95}
\bibinfo{author}{\bibfnamefont{D.~M.} \bibnamefont{Ceperley}},
  \bibinfo{journal}{Rev. Mod. Phys.} \textbf{\bibinfo{volume}{67}},
  \bibinfo{pages}{279} (\bibinfo{year}{1995}).

\bibitem[{\citenamefont{Ceperley}(1991)}]{ceperley1991fermion}
\bibinfo{author}{\bibfnamefont{D.~M.} \bibnamefont{Ceperley}},
  \bibinfo{journal}{J. Stat. Phys.} \textbf{\bibinfo{volume}{63}},
  \bibinfo{pages}{1237} (\bibinfo{year}{1991}).

\bibitem[{\citenamefont{Weiss and Egger}(2005)}]{PIMC05}
\bibinfo{author}{\bibfnamefont{S.}~\bibnamefont{Weiss}} \bibnamefont{and}
  \bibinfo{author}{\bibfnamefont{R.}~\bibnamefont{Egger}},
  \bibinfo{journal}{Phys. Rev. B} \textbf{\bibinfo{volume}{72}},
  \bibinfo{pages}{245301} (\bibinfo{year}{2005}).

\bibitem[{\citenamefont{Boninsegni et~al.}(2006)\citenamefont{Boninsegni,
  Prokof'ev, and Svistunov}}]{boninsegni2006worm}
\bibinfo{author}{\bibfnamefont{M.}~\bibnamefont{Boninsegni}},
  \bibinfo{author}{\bibfnamefont{N.}~\bibnamefont{Prokof'ev}},
  \bibnamefont{and}
  \bibinfo{author}{\bibfnamefont{B.}~\bibnamefont{Svistunov}},
  \bibinfo{journal}{Phys. Rev. Lett.} \textbf{\bibinfo{volume}{96}},
  \bibinfo{pages}{070601} (\bibinfo{year}{2006}).

\bibitem[{\citenamefont{Yan and Blume}(2013)}]{PIMC13}
\bibinfo{author}{\bibfnamefont{Y.}~\bibnamefont{Yan}} \bibnamefont{and}
  \bibinfo{author}{\bibfnamefont{D.}~\bibnamefont{Blume}},
  \bibinfo{journal}{Phys. Rev. A} \textbf{\bibinfo{volume}{88}},
  \bibinfo{pages}{023616} (\bibinfo{year}{2013}).

\bibitem[{\citenamefont{Yan and Blume}(2014)}]{PIMC14}
\bibinfo{author}{\bibfnamefont{Y.}~\bibnamefont{Yan}} \bibnamefont{and}
  \bibinfo{author}{\bibfnamefont{D.}~\bibnamefont{Blume}},
  \bibinfo{journal}{Phys. Rev. Lett.} \textbf{\bibinfo{volume}{112}},
  \bibinfo{pages}{235301} (\bibinfo{year}{2014}).

\bibitem[{\citenamefont{Dornheim et~al.}(2018)\citenamefont{Dornheim, Groth,
  and Bonitz}}]{dornheim2018uniform}
\bibinfo{author}{\bibfnamefont{T.}~\bibnamefont{Dornheim}},
  \bibinfo{author}{\bibfnamefont{S.}~\bibnamefont{Groth}}, \bibnamefont{and}
  \bibinfo{author}{\bibfnamefont{M.}~\bibnamefont{Bonitz}},
  \bibinfo{journal}{Phys. Rep.} \textbf{\bibinfo{volume}{744}},
  \bibinfo{pages}{1} (\bibinfo{year}{2018}).

\bibitem[{\citenamefont{Cui}(2012)}]{Cui12}
\bibinfo{author}{\bibfnamefont{X.}~\bibnamefont{Cui}}, \bibinfo{journal}{Phys.
  Rev. A} \textbf{\bibinfo{volume}{85}}, \bibinfo{pages}{022705}
  (\bibinfo{year}{2012}).

\end{thebibliography}

\end{document}